\documentclass[a4paper,USenglish,cleveref,autoref,thm-restate]{lipics-v2021}       

\title{Characterizing Feedback Statements in Machine Learning Jupyter Notebooks}

\author{Arumoy Shome}{Delft University of Technology, The Netherlands}{a.shome@tudelft.nl}{https://orcid.org/0000-0002-3778-5150}{}
\author{Lu\'is Cruz}{Delft University of Technology, The Netherlands}{l.cruz@tudelft.nl}{https://orcid.org/0000-0002-1615-355X}{}
\author{Diomidis Spinellis}{Delft University of Technology, The Netherlands}{d.spinellis@tudelft.nl}{https://orcid.org/0000-0003-4231-1897}{}
\author{Arie van Deursen}{Delft University of Technology, The Netherlands}{arie.vandeursen@tudelft.nl}{https://orcid.org/0000-0003-4850-3312}{}

 \authorrunning{A. Shome, L. Cruz, D. Spinellis, and A.\,V. Deursen} 

\Copyright{Arumoy Shome, Lu\'is Cruz, Diomidis Spinellis, and Arie van Deursen} 



\EventEditors{Robert Feldt, Maria Paasivaara, Daniel Mendez, Stefan Wagner, and Marvin Mu\~{n}oz Bar\'{o}n}
\EventNoEds{5}
\EventLongTitle{20th International Symposium on Empirical Software Engineering and Measurement (ESEM 2026)}
\EventShortTitle{ESEM 2026}
\EventAcronym{ESEM}
\EventYear{2026}
\EventDate{October 8--9, 2026}
\EventLocation{Munich, Germany}
\EventLogo{}
\SeriesVolume{394}
\ArticleNo{13}

\usepackage{hyperref}
\usepackage{cite}
\usepackage{amsmath,amssymb,amsfonts}
\usepackage{algorithmic}
\usepackage{graphicx}
\usepackage{booktabs}
\usepackage{textcomp}
\usepackage{xcolor}
\usepackage{tabularx}
\usepackage[most]{tcolorbox}
\usepackage{listings}
\usepackage{subcaption}
\usepackage{longtable}
\usepackage{framed}

\definecolor{backcolor}{rgb}{0.95,0.95,0.92}
\colorlet{punct}{red!60!black}
\definecolor{delim}{RGB}{20,105,176}
\colorlet{numb}{magenta!60!black}

\lstdefinestyle{mystyle}{
  language=Python,
  basicstyle=\ttfamily\footnotesize,
  keywordstyle=\color{delim},
  stringstyle=\color{punct},
  commentstyle=\color{numb},
  breakatwhitespace=true,
  breaklines=true,
  keepspaces=true,
  showspaces=false,
  showstringspaces=false,
  showtabs=false,
  tabsize=2,
  stepnumber=1,
  numbersep=3pt,
}
\nolinenumbers

\begin{document}

\maketitle

\begin{abstract}
\textbf{Background.}
Machine learning development in Jupyter notebooks
is iterative and feedback-driven.
Practitioners author statements
that reveal information about program execution
and use this information
to decide what to do next.
We call these \emph{feedback statements}
and identify two forms:
exploratory statements that display values for visual inspection
and validation statements that enforce conditions programmatically through assertions.

\textbf{Aims.}
Many failures in ML systems do not surface as exceptions,
and consequently escape the crash-based analyses
that dominate prior empirical work on ML notebooks.
This study examines what practitioners check
to catch the failures that would otherwise pass silently,
by characterizing feedback statements
that encode the practitioner's mental model
of what the code should do
and what could go wrong.

\textbf{Method.}
We mine 297,851 publicly available Python Jupyter notebooks
from GitHub and Kaggle,
and extract 1,092,780 feedback statements.
We sample 816 statements
through proportional stratified sampling
from semantic clusters obtained from CodeBERT embeddings,
and apply grounded theory and open coding
to manually label and analyze each statement.

\textbf{Results}.
We contribute a taxonomy of feedback statements in ML Jupyter notebooks,
organized along the functional intent of the statement
and the ML pipeline stage in which it appears.
The taxonomy reveals
that feedback in ML notebooks is overwhelmingly exploratory,
and that the two platforms host qualitatively different modes of ML work.
We further map our taxonomy to an existing crash taxonomy
and find that our taxonomy captures defensive practices against silent failures
that crash analysis cannot observe.

\textbf{Conclusions.}
Our findings indicate
that notebook source should be treated
as a confounder in studies of ML developer practice,
surface opportunities for notebook tooling,
and motivate empirical study of silent ML failures.
We release the corpus of 1,092,780 feedback statements and the codebook,
to support replication and tooling research.
\end{abstract}

\section{Introduction}

Machine learning development is iterative and exploratory.
Practitioners assemble data from heterogeneous sources,
clean and preprocess it,
build candidate models,
evaluate them,
revise feature engineering choices,
and repeat the cycle until the model
satisfies the target criteria~\cite{haakman2021ai,amershi2019software,sculley2015hidden}.
Each cycle is driven by feedback,
closely resembling prominent development processes,
such as Agile~\cite{betz2018managing} and CRISP-DM~\cite{martinez-plumed2021crisp-dm},
that advocate for short feedback loops
and frequent revision in response to what each cycle reveals.

Jupyter notebooks have become the dominant medium for developing ML prototypes
since they support such a feedback-driven workflow,
through a \emph{read-eval-print} loop style of development.
Practitioners write a small unit of code within a \emph{cell},
execute it,
read the result,
and decide what to do next~\cite{pimentel2019large-scale,quaranta2021kgtorrent,psallidas2019data,perkel2018why}.
The output of each cell is thus a unit of feedback,
that is made visible by the notebook structure,
and can be traced back to a particular code statement in the cell.

Output can originate from statements that contain a bug,
and crash when executed.
Existing empirical work on ML notebooks
has examined such points of failure,
by analyzing crashes~\cite{wang2025why}, bugs~\cite{desantana2024bug} and vulnerabilities~\cite{jiang2025exploring}
that occur within Jupyter notebooks.
However, many failures in ML systems do not surface as exceptions.
Sculley et al.~\cite{sculley2015hidden} show
that machine learning carries a higher technical debt burden than conventional software,
since data dependencies, feedback loops, and configuration changes
can degrade model behavior without raising an error.
Breck et al.~\cite{breck2019data} make a similar argument for training data,
and show that a single corrupted feature
can shift a model's predictions
while the training pipeline continues to execute without complaint.

In contrast to prior work,
this study focuses on \emph{feedback statements},
which we define as statements that the practitioner intentionally adds to the code
to obtain information about the program's execution.
Practitioners rely on this feedback
to confirm that the program behaves as intended,
and to catch the failures that would otherwise pass silently.
Based on this definition,
code statements that crash and produce an output,
are excluded from this study.
Although a bug that produces an error message also provides feedback,
it does not satisfy our definition
because the practitioner did not author the bug to obtain feedback.

We identify two forms of feedback statements that satisfy our definition.
First, we study exploratory feedback (EXP)
obtained from the last statement in a code cell,
that display values for the practitioner to interpret.
Our decision is grounded in prior studies~\cite{kery2018story,rule2018exploration}
that have established such statements
to be the primary mode of exploratory programming in Jupyter notebooks.
Second, we study validation statements (VAL)
that enforce a condition programmatically,
and halt execution with an explicit signal
when this condition is violated.
This form of feedback is obtained from \lstinline{assert} statements,
which are used by practitioners to validate assumptions regarding a piece of code.
Notebooks are primarily used to develop a prototype of an ML model,
prior to operationalizing it within a production environment~\cite{shankar2024we,nahar2023meta-summary}.
Assertions are therefore a suitable candidate for studying validation behavior,
since they are commonly employed in the early stages of the development lifecycle~\cite{kochhar2017revisiting}.
Pimentel et al.~\cite{pimentel2019large-scale} report
that unit tests are virtually absent from Jupyter notebooks,
but since Python unit tests are themselves composed of one or more assert statements,
our approach does not miss them.

We mine 297,851 publicly available Python Jupyter notebooks
from GitHub and Kaggle,
and extract 1,092,780 feedback statements.
We sample 816 statements
using proportional stratified sampling
from semantic clusters obtained from CodeBERT embeddings.
We use grounded theory
and analyze each statement manually along two dimensions:
\textbf{(1)} the \emph{intent} of the statement,
which captures what the practitioner is trying to learn or confirm,
and \textbf{(2)} the ML \emph{pipeline stage},
where the statement appears.
We use a combination of statistical hypothesis testing
and qualitative analysis
to answer the following research questions.

\begin{description}
  \item[RQ1.] \textbf{What do practitioners intend to learn or confirm from feedback statements?}

    We find that
    practitioners predominantly inspect, compute, and probe structure when exploring,
    and check shape, equality, and boolean invariants when validating.

  \item[RQ2.] \textbf{How prevalent are exploratory and validation feedback statements in ML notebooks?}

    Feedback in ML notebooks is overwhelmingly exploratory,
    and account for 85\% of the corpus.
    Formal assertions account for the remaining 15\%
    and are concentrated almost entirely in GitHub.

  \item[RQ3.] \textbf{Where in the ML pipeline do practitioners author feedback statements?}

    The two platforms host qualitatively different modes of ML work.
    Kaggle notebooks concentrate 58\% of feedback in the data preparation stage
    and barely touches the model construction code.
    GitHub notebooks on the other hand,
    distribute feedback across data ingestion, model construction, training, and evaluation.

  \item[RQ4.] \textbf{How does the feedback taxonomy relate to known crash modes in ML notebooks?}

    We establish a mapping between our feedback taxonomy
    and the crash taxonomy of Wang et al.~\cite{wang2025why},
    and find that both converge on data preparation.
    Our taxonomy additionally captures defensive practices
    that prevent silent failures,
    which are otherwise invisible to crash analysis.
\end{description}

We identify a repertoire of exploratory feedback statements
that practitioners author on both platforms,
dominated by inspection and exploratory computation.
We find wide use of validation on GitHub,
especially shape and equality checks,
but near absence on Kaggle.
Kaggle further concentrates feedback in data preparation,
while GitHub distributes it across all pipeline stages.
Our findings indicate that notebook source
should be treated as a confounder in studies of ML developer practice,
surface opportunities for notebook tooling
that builds on what practitioners already author,
and motivate empirical study of silent ML failures
that crash-based analysis cannot observe.
We release a publicly available dataset
of 1M+ feedback statements
together with the codebook,
to support replication and downstream tooling research~\cite{replication}.

\section{Methodology}
\label{sec:method}

\begin{figure}
 \includegraphics[width=\textwidth]{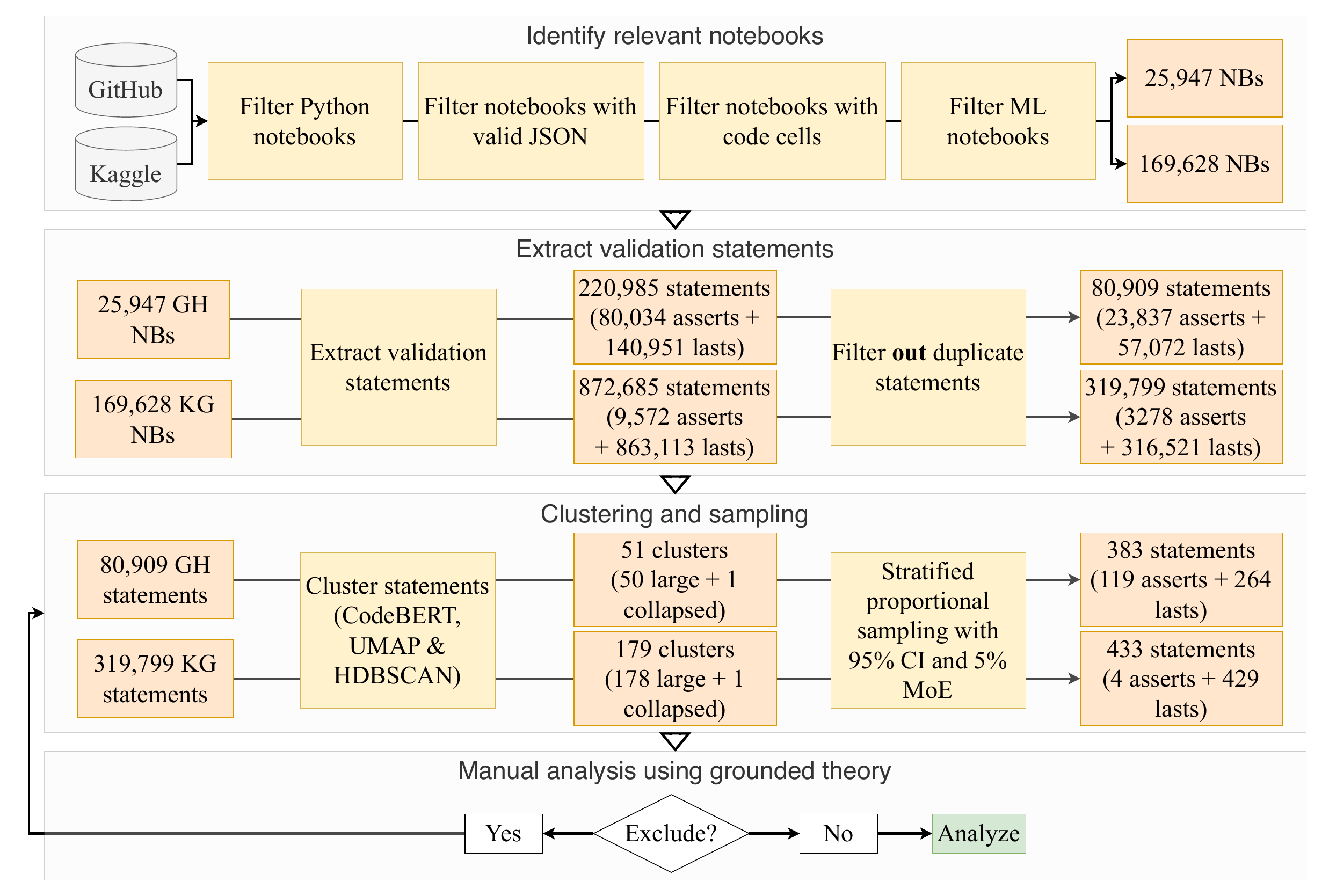}
 \caption{Overview of the data collection, data filtering, clustering, sampling, and manual analysis
 of feedback statements
from GitHub and Kaggle.}\label{fig:method}
\end{figure}

Using the Goal Question Metric Paradigm~\cite{caldiera1994goal},
the research goal of this study
is to \emph{characterize}
the feedback statements used in ML Jupyter notebooks,
based on their intent
and the ML pipeline stage in which they occur.
This research goal guides the data collection and analysis
methodology outlined in this section.

Figure~\ref{fig:method} illustrates the
data collection, filtering, sampling, and manual analysis
employed in this study
to gather and analyze feedback statements
written in Python ML notebooks.
We collect feedback statements from GitHub (GH)
that hosts general software projects under version control,
and Kaggle (KG), which is focused on data science and machine learning competitions.

We use GitHub's advanced search syntax~\footnote{https://docs.github.com/en/search-github/searching-on-github/searching-code},
\footnote{We use the following search query: \lstinline[language={}]$language:"Jupyter Notebook"$}
to collect \emph{all} Jupyter notebooks from public repositories
authored on June 22, 2023.
This is because mining software repositories on GitHub
based on popularity metrics (such as stars, forks, pull requests, etc.)
are largely contested in the MSR community~\cite{kalliamvakou2015in-depth}.
Our approach is also grounded in prior work~\cite{jiang2022elevating}
that adopt a similar strategy to mine ML Jupyter notebook from GitHub.
We use the pre-existing dataset KGTorrent~\cite{quaranta2021kgtorrent,kgtorrent-zenodo} for Kaggle
since it does not support advanced code-based search like GitHub.
Consequently, we start with 297,851 notebooks (283 GB)
comprising 49,090 notebooks from GitHub and 248,761 notebooks from Kaggle.

\textbf{Extracting validation statements}.
Since the focus of this study is to analyze Python code in ML Jupyter notebooks,
we exclude notebooks that are not authored in the Python programming language,
contain invalid JSON structures,
or no code cells.
To concentrate specifically on machine learning projects,
we only include notebooks that import at least one major ML library:
Scikit Learn~\cite{pedregosa2011scikit-learn},
PyTorch~\cite{paszke2017automatic},
TensorFlow~\cite{abadi2015tensorflow},
or Keras~\cite{chollet2015keras}.
While this approach may exclude some
ML projects using alternative libraries,
Psallidas et al.~\cite{psallidas2019data}
examined six million GitHub Jupyter notebooks
and confirmed that most ML-oriented notebooks utilize these frameworks.
After applying these
filters, we exclude 102,276 (41\%) notebooks
and retain 195,575 notebooks (25,947 from GH and 169,628 from KG) for analysis.

We extract the contents of \emph{all} code cells from the notebooks
by parsing the underlying JSON structure,
and then collect \emph{all} assert and last statements from these cells
using Python's \lstinline{ast} module.
For assertions, we gather both built-in \lstinline{assert} statements
and assertion-style calls from third-party libraries
(i.e., function calls whose name contains ``assert''),
which lets us capture validation behavior beyond the Python standard library.
For last statements, we restrict our analysis to cells that produce an output
and extract the final top-level statement from each.
Consequently, we obtain a population of
220,095 statements (80,034 asserts and 140,951 lasts) from GitHub
and 872,685 statements (9,572 asserts and 863,113 lasts) from Kaggle.

\textbf{Clustering and sampling.}
In a preliminary analysis,
we apply clustering to the entire dataset,
and manually analyze 100 statements sampled randomly from GH and KG.
We find that duplicate statements conflate the cluster structure
and cause a random sample to under-represent diverse patterns.
We therefore remove exact duplicate statements prior to clustering,
and deliberately trade off diversity with representativeness for the taxonomy.

We use CodeBERT to obtain semantic code embeddings for each unique statement.
Semantic embeddings are appropriate
because statements with different surface forms can express the same intent.
E.g., \lstinline{assert X == Y} and \lstinline{assert X is Y}
are syntactically distinct but semantically related.
We use UMAP~\cite{mcinnes2018} to reduce the embedding dimensionality to 10,
and apply HDBSCAN~\cite{mcinnes2017} to cluster the reduced embeddings
based on their semantic similarity.
We obtain 277 clusters from GH and 1031 clusters from KG.
Prior to sampling,
we collapse small clusters to prevent excessive fragmentation of the strata.
We rank clusters by size
and compute their cumulative coverage of the entire population.
Based on prior work~\cite{wang2025why},
clusters that together account for the top 70\% of statements
are retained as individual strata.
All remaining clusters are merged into a single residual stratum,
which ensures that each retained stratum
is large enough to yield a meaningful proportional allocation.

We apply proportional stratified sampling
with a 95\% confidence interval (CI)
and a 5\% margin of error (MoE)
to estimate a representative sample of feedback statements.
This gives a minimum sample size of 383 statements for GH
and 384 statements for KG.
To ensure comprehensive coverage,
we sample from \emph{all} strata
and ensure that at least one statement is sampled from each stratum.
In total, we obtain 383 statements from GitHub
and 433 statements from Kaggle.

\textbf{Data Analysis}.
We apply grounded theory~\cite{seaman1999qualitative}
and analyze all feedback statements manually.
A codebook is developed using a seed of 50 statements,
from the top 5 clusters.
Two members of the research team analyze each statement independently,
to answer ``What the developer is trying to learn or confirm?'' with the statement,
and ``In which stage of the ML application development pipeline?'' is the statement defined.
In line with Amershi et al.~\cite{amershi2019software},
we distinguish six stages,
corresponding to data collection (DATA),
preparation (PREP),
feature engineering (FEAT),
model construction (MODEL),
training (TRAIN),
and evaluation (EVAL).

In addition to the statement,
the entire code cell of the statement,
any related code cells where variables or functions are defined,
and the markdown cells containing documentation is analyzed to bring in additional context.
The authors then meet to compare the codes and discuss on disagreements
until consensus is reached.
After four additional iterations ($50 \times 5=250$ statements),
saturation is observed
as no new codes are added to the codebook.
The first author then analyzes and codes the remaining 566 statements.
To verify the coding done by the first author on the 566 statements,
the second author independently analyzes 125 (22\%) of these statements,
and a Cohen's $\kappa$ score of 0.93 for the intent dimension,
and 0.89 for the stage dimension is observed,
which indicates high agreement.

While conducting the manual analysis,
we observe 90 feedback statements that fall outside the scope of this study,
grouped into 6 exclusion categories.
We exclude exploratory statements
used to probe an API
or inspect how an imported function behaves;
statements from scientific simulations
that do not involve training an ML model;
misused code cells
that contain block-quoted documentation
or commented-out code with stale output;
statements with no functional intent relevant to this study,
such as \lstinline{plt.ylabel('Win 
statements not authored in English;
and assertions found in academic notebooks used for automatic grading.
To preserve our target sample size,
we resample replacements from the same clusters where the exclusions occurred.
Replacements are verified
and if necessary replaced once again,
until the target sample size is reached.

\textbf{Statistical framework}.
We test whether two categorical variables are associated,
with the null hypothesis
that the variables are independent.
We adopt a 95\% confidence level
and treat any p-value below 0.05
as a statistically significant association.
We select the test
based on the expected cell frequencies of the contingency table.
When all expected frequencies are at least five
and Cochran's rule holds,
we apply Pearson's chi-squared ($\chi^2$) test.
Otherwise we apply Fisher's exact test
with Monte Carlo simulation ($n=9999$ resamples).
We also report the bias-corrected Cram\'er's $V$ as the effect size for all tests
(small: $V < 0.1$, medium: $0.1 \le V < 0.3$, large: $V \ge 0.3$).
Before each test we verify that the validity conditions hold
and we report the outcome by analyzing the residuals that drive the association.

For RQ2 and RQ3,
we apply family-wise Bonferroni correction,
and adopt an adjusted threshold of $\alpha = 0.05/2 = 0.025$
since each RQ contains a family of two tests.


When the marginal distribution of a variable is severely imbalanced,
no test has adequate power and we report descriptive statistics only.

\section{Results}

\begin{table}
  \centering
  \caption{Taxonomy of feedback statements in ML Jupyter notebooks.
  The intent dimension distinguishes exploratory feedback (EXP)
  from formal validation (VAL).
  Each intent is refined into type subcodes
  derived through open coding of 816 feedback statements.
  Percentage is normalized within each top-level intent.}\label{tab:taxonomy}
  \begin{tabularx}{\textwidth}{l p{2cm} X X r}
    \toprule
    \multicolumn{5}{l}{\emph{\textbf{Validation Statements (VAL)}}} \\
    \midrule
    \textbf{Type} &
    \textbf{Label} &
    \textbf{Definition} &
    \textbf{Example} &
    \textbf{$n$(\%)} \\
    \midrule
    SHAPE &
    Dimensionality check &
    Assert tensor/array shape or length &
    \lstinline{assert x.shape == means.shape} &
    47 (39\%) \\
    EQ &
    Value equality check &
    Assert that a result equals a known correct value &
    \lstinline{assert_frame_equal(...)} &
    35 (28\%) \\
    BOOL &
    Boolean invariant &
    Assert a boolean condition holds &
    \lstinline{assert bb2['y1'] < bb2['y2']} &
    21 (17\%) \\
    APPROX &
    Approximate equality &
    Assert numerical closeness within tolerance &
    \lstinline{assert np.allclose(..., atol=0.001)} &
    8 (6\%) \\
    EXIST &
    Check for existence of value &
    Assert that a value exists &
    \lstinline{assert i in sig.parameters.keys()} &
    6 (5\%) \\
    TYPE &
    Type check &
    Assert object type or class membership &
    \lstinline{assert type(embedding_layer) == Embedding} &
    6 (5\%) \\
    \midrule

    \multicolumn{5}{l}{\emph{\textbf{Exploratory Statements (EXP)}}} \\
    \midrule
    INSPECT &
    Object inspection &
    Display and visually verify an object &
    \lstinline{df.head()} &
    303 (44\%) \\
    COMPUTE &
    Exploratory computation &
    Inspect the result of a transformation &
    \lstinline{.sort_values()} &
    186 (27\%) \\
    STRUCT &
    Structural probe &
    Inspect shape, size, keys, columns and type metadata &
    \lstinline{df.shape} &
    126 (18\%) \\
    STATS &
    Statistical summary &
    Display descriptive statistics &
    \lstinline{df.describe()} &
    78 (11\%) \\
    \bottomrule
  \end{tabularx}
\end{table}

This section presents our findings for the four research questions.
In Section~\ref{sec:rq1}, we present the taxonomy of feedback statements
derived from open coding,
and describe the subtypes within each top-level intent.
Section~\ref{sec:rq2} establishes what feedback statements practitioners author
and how their prevalence varies between GitHub and Kaggle.
Section~\ref{sec:rq3}, presents a cross-cutting analysis with the ML pipeline stage
to characterize where each kind of feedback appears
in the development workflow.
Finally, in Section~\ref{sec:rq4}
we relate the feedback taxonomy to the crash taxonomy from Wang et al.~\cite{wang2025why}
to surface the duality between defensive practice and observed failure.
A summary of the feedback taxonomy is presented in Table~\ref{tab:taxonomy}.

\subsection{RQ1: Intent Taxonomy of Feedback Statements}
\label{sec:rq1}

Open coding of the 816 feedback statements
produces ten types
distributed across the two top-level intents
(see Table~\ref{tab:taxonomy}).
We briefly describe each type below
and ground the description in the use cases that emerge from the data.
We organize the description by the frequency of the intent subtypes.
A more detailed version of the taxonomy
along with axial and boundary coding decisions for the codebook
can be found in our replication package~\cite{replication}.

\begin{lstlisting}[caption={Check that the shape of the fifth feature map from a VGG16 encoder
immediately before it is connected to a segmentation decoder.}, label={lst:61}]
assert output['x5'].size() == torch.Size([BATCH_SIZE, 512, 7, 7])
\end{lstlisting}

\textbf{Dimensionality check (VAL-SHAPE)}.
Shape validation is the most prevalent form of explicit validation in the corpus.
Practitioners verify
that arrays and tensors carry the dimensions
expected by downstream operations,
since a shape mismatch either raises a cryptic runtime error far from its origin,
or silently produces wrong results through unintended broadcasting.

In the upstream data stages,
practitioners assert that a train-test split accounts for all rows,
that a join did not silently add or drop records,
and that a preprocessing step such as \lstinline{QuantileTransformer} preserves the sample count.
At the model construction stage,
practitioners verify that each layer of a neural network produces the expected output shape
before the next layer is wired in (Listing~\ref{lst:61}).
At the evaluation stage,
shape validation guards against
computing metrics (e.g., RMSE or F1)
on misaligned prediction and label arrays,
which either crashes the metric computation
or produces a numerically plausible but incorrect score.

\begin{lstlisting}[caption={Check that the sinusoidal positional embeddings have an even dimension
because the vector is split between sine and cosine components.}, label={lst:4}]
assert dim % 2 == 0
\end{lstlisting}

\textbf{Value equality check (VAL-EQ)}.
Practitioners check that a specific value, configuration parameter, or computed result
matches an expected value,
which is known at the time of writing the assertion
and is derived from domain knowledge, architectural constraints, or a prior specification.

Practitioners perform schema validation on loaded data
by checking that a CSV file has the expected number of classes,
that a required column appears in the correct position,
that an index contains no duplicates,
and that a normalization step produces the expected result.
Assertions also enforce model architectural constraints
that the framework does not check at runtime.
For instance, without the assertion in Listing~\ref{lst:4},
a violated constraint would surface as a shape error deep inside the forward pass,
whose connection to the misconfigured hyperparameter is not obvious.
A third use is the detection of training-serving skew (Listing~\ref{lst:118}),
where a divergence between the offline and production preprocessing pipelines
would cause the deployed model
to operate on a different feature distribution than the one it was trained on,
producing a silent failure with no crash signal.

\begin{lstlisting}[caption={Verify that the offline preprocessing pipeline
and the production preprocessing pipeline
produce identical training data after feature selection.}, label={lst:118}]
assert_frame_equal(X_train_prepro_ks.to_pandas(), X_train_prod_prepro_ks.to_pandas())
\end{lstlisting}

\textbf{Boolean invariant (VAL-BOOL)}.
Boolean validation captures a wide range of preconditions
that do not reduce to a shape check or an exact equality.
The most common theme
is to enforce data quality checks
before a downstream step
that cannot recover from a violated precondition.
Practitioners assert the absence of missing values before feature selection,
the absence of class imbalance before oversampling logic executes,
and that domain constraints are satisfied
before domain-specific computations begin (Listing~\ref{lst:54}).

\begin{lstlisting}[caption={This assertion appears in a cardiovascular disease mortality pipeline
to prevent missing values from silently corrupting feature selection
or crashing estimators that assume complete data.}, label={lst:54}]
assert not predictors.isnull().values.any()
\end{lstlisting}

\textbf{Approximate equality (VAL-APPROX)}.
Approximate validation appears where exact equality is either mathematically inappropriate
or not possible due to floating-point arithmetic.
Practitioners validate custom loss functions,
data transformation utilities,
and metric implementations
by comparing their output to a known reference value
within a specified tolerance (Listing~\ref{lst:374}).

\begin{lstlisting}[caption={Validate that a custom NDCG implementation
returns a value close to 1.0
when predicted and true rankings are identical.}, label={lst:374}]
assert np.isclose(ndcg([3, 1, 2], [3, 1, 2], 3), 1)
\end{lstlisting}

\textbf{Existence check (VAL-EXIST)}.
Existence validation is used to prevent data leakage
between the validation and training sets (Listing~\ref{lst:84}),
which inflates evaluation metrics without any runtime error.
A second theme is precondition checking on external resources.
Practitioners assert that a required file path exists before attempting to load it,
and that a hyperparameter name is present in the parameter space of an estimator
before launching an expensive grid search.
In both cases the failure mode without the assertion
is either a cryptic I/O error
or a silent search over the wrong parameter space.

\begin{lstlisting}[caption={Verifies that training and validation sets are disjoint
before model fitting begins.}, label={lst:84}]
assert set(train_dataset.filenames).isdisjoint(set(valid_dataset.filenames))
\end{lstlisting}

\textbf{Type check (VAL-TYPE)}.
Practitioners use these statements to enforce
that objects carry the correct Python type,
before passing them to operations that would fail silently
or produce a misleading error if given the wrong type.
A common pattern asserts that model parameters are NumPy arrays rather than Python lists,
since many NumPy operations accept both
but produce different results.
The low frequency of VAL-TYPE is likely due to
Python's dynamic typing and framework-level type coercion
which often suppress type errors silently.

\textbf{Object inspection (EXP-INSPECT)}.
Inspection statements are the most frequent form of exploratory feedback in the corpus.
A practitioner places a bare variable name,
a plot call,
or an accessor expression
as the last statement in a cell
to visually check the current state of a data structure.

Inspection is used throughout the pipeline
but most prominent at the data loading and preparation stages,
where practitioners establish a running mental model of the dataset
before any transformation is applied.
A bare \lstinline{.head()} call after loading a CSV
confirms that column names were parsed correctly,
and that data types look plausible.
The same lightweight pattern monitors the effect of individual cleaning operations,
such as inspecting a column after imputing missing values to confirm that the fill worked.
During training,
a recurring pattern is the periodic visual inspection of generated outputs
or loss curves.
At the evaluation stage,
inspection shifts from data structure verification to model output monitoring.
Practitioners surface intermediate predictions,
confusion matrices,
learned coefficients,
and cross-validation results as bare last-cell expressions.

A theme distinctive to GitHub notebooks
is the use of inspection to debug internal model state.
Practitioners surface the names of specific layers in a neural network,
the values of learned weight slices,
or the output of a forward hook
to understand what the model is doing internally.

\textbf{Exploratory computation (EXP-COMPUTE)}.
Compute statements execute a transformation, aggregation, or model operation
and return a result that the practitioner examines.
The distinction from inspection is that the result is not already present in a variable
but is produced on the fly by the statement itself.

The dominant use of EXP-COMPUTE in both GitHub and Kaggle notebooks
is metric computation at the evaluation stage.
Practitioners compute accuracy, RMSE, F1, NDCG, AUC, and log-loss
as last-cell expressions
to obtain an immediate reading of model performance.
These calls are repeated across multiple experimental runs
to drive an iterative model and hyperparameter selection process.

A second theme is informal postcondition checking on preprocessing steps.
After applying a standard scaler,
practitioners compute the mean and standard deviation of the transformed features
to verify that both are near their expected values.
Similarly, after a \lstinline{dropna} operation,
practitioners compute a null count to confirm that missing values were removed.

In Kaggle notebooks,
EXP-COMPUTE is used extensively for exploratory feature analysis.
Practitioners compute grouped aggregations and correlation matrices
to understand whether a candidate feature has a meaningful relationship with the target
before deciding whether to include it.

\textbf{Structural probe (EXP-STRUCT)}.
Structural inspection statements surface
shape, column names, dtypes, unique values, cardinality, or value counts of a data structure.
A pervasive theme is to use structural probes
to inform how a data transformation should be configured,
and after to confirm that it had the intended effect.
For instance, practitioners use \lstinline{.unique()} and \lstinline{.value_counts()}
to audit the actual contents of a categorical column before deciding on an encoding strategy,
and check \lstinline{.shape} after one-hot encoding
to confirm that the column count increased as expected.

\textbf{Statistical summary (EXP-STATS)}.
Practitioners use distribution plots and summary statistics
to detect outliers,
assess class imbalance,
and check whether a feature's distribution motivates a transformation.
Similar to structural probes,
these statements appear on both sides of a transformation.
For instance, after scaling features with \lstinline{StandardScaler},
practitioners call \lstinline{.describe()}
to confirm that the mean is near zero and the standard deviation is near one.

\subsection{RQ2: Prevalence and Variation of Feedback Statements}
\label{sec:rq2}

\begin{figure}
  \begin{subfigure}[t]{0.4\textwidth}
    \centering
    \includegraphics[width=\linewidth]{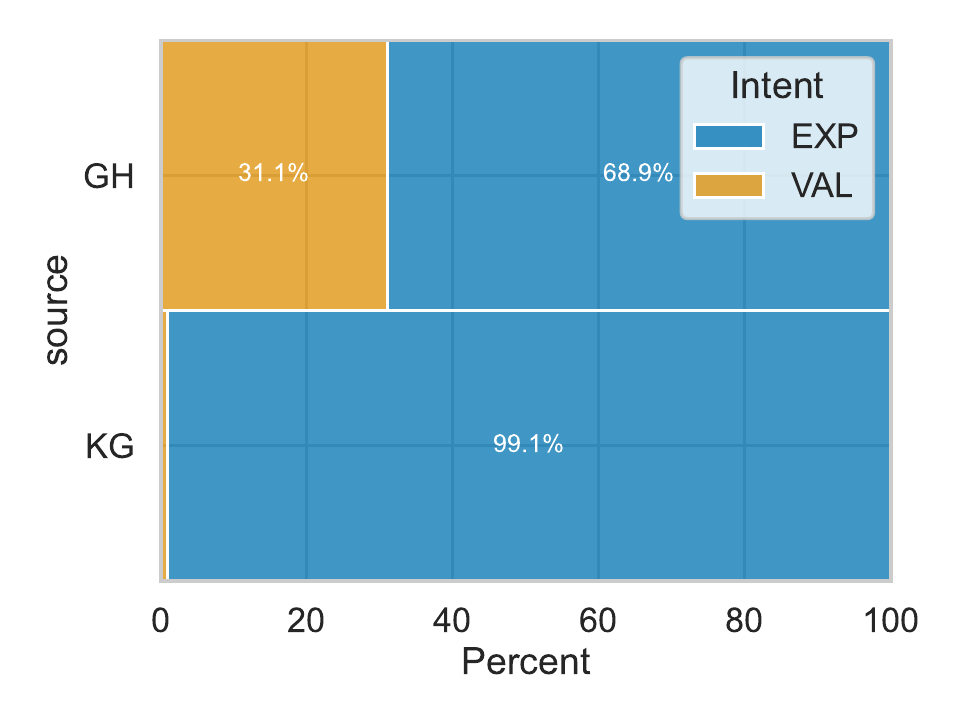}
    \caption{Top-level intent (EXP vs.\ VAL) across GH and KG, normalized within source.}
    \label{fig:distribution-intent}
  \end{subfigure}
  \begin{subfigure}[t]{0.6\textwidth}
    \centering
    \includegraphics[width=\linewidth]{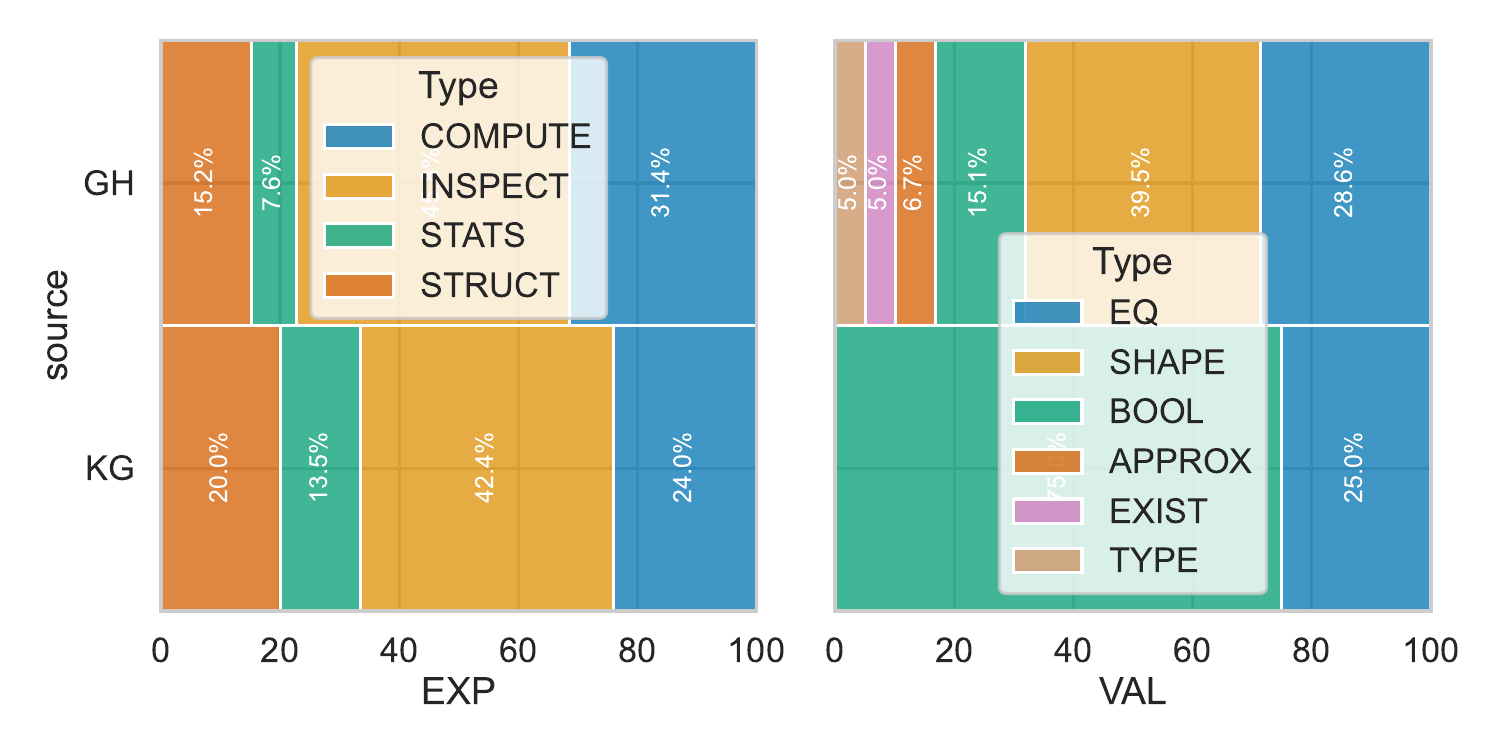}
    \caption{Subtype distribution within each top-level intent, normalized within intent.}
    \label{fig:distribution-type}
  \end{subfigure}
  \caption{Distribution of feedback statements across top-level intent and its subtypes.
  GitHub contains a substantially higher share of formal assertions than Kaggle.
  Within EXP statements, the INSPECT subtype dominates on both platforms.
  Within VAL statements, SHAPE and EQ subtypes together account for two-thirds of all assertions.}
\end{figure}

\textbf{Feedback in ML notebooks is overwhelmingly exploratory.}
Exploratory statements (EXP) account for 85\% ($n=693$) of the corpus
and validation statements (VAL) account for the remaining 15\% ($n=123$).
The split is highly skewed toward exploratory feedback on both platforms.

We hypothesise that the intent of the feedback statements
on GitHub and Kaggle differ based on the contrasting purposes of the two platforms.
GitHub hosts software engineering projects
where assertions are a natural practice,
while Kaggle hosts data exploration and competition submission workflows
where programmatic validation is uncommon.
A $\chi^2$ test between intent and notebook source
reveals a significant and large association ($\chi^2(1) = 144.72$, $p < 0.001$, Cram\'er's $V = 0.421$),
and the direction of the association matches the prediction.
Figure~\ref{fig:distribution-intent} shows that
GH notebooks contain proportionally more formal assertions (31\%),
while KG notebooks are overwhelmingly exploratory
and contain only four assertions (1\%).
Section~\ref{sec:rq3} explores this asymmetry further,
where we show that the source interacts with pipeline stage as well.

For EXP statements ($n=693$),
a $\chi^2$ test reveals a statistically significant but weak association
between type and source
($\chi^2(3) = 10.77$, $p=0.013$, Cram\'er's $V=0.125$),
further confirming that the two platforms
are used for different purposes.
For VAL statements ($n=123$),
the marginal imbalance leaves too little power for a meaningful test of independence.
Figure~\ref{fig:distribution-type} therefore presents
the type distribution descriptively for both intents.

Within EXP statements,
practitioners most often display values
to confirm what an expression evaluates to (intent type coded as INSPECT),
compute derived quantities they read off the cell output (COMPUTE),
probe the structure of an object (STRUCT),
and least often request a statistical summary (STATS).
This rank order is consistent with a \emph{notebook-as-scratchpad} model of feedback,
and holds on both platforms.
Within VAL statements,
practitioners predominantly assert dimensional (SHAPE) and equality (EQ) invariants.
Type checks, existence checks, and approximate equality are present
but appear in fewer than one in ten validation statements.

\textbf{Of the 123 VAL statements, only four occur in Kaggle notebooks.}
Within these assertions, four of the six VAL types (SHAPE, APPROX, EXIST, TYPE)
are entirely absent,
suggesting that Kaggle contributors rarely use structural validation.

\subsection{RQ3: Pipeline Stage Distribution of Feedback Statements}
\label{sec:rq3}

\begin{figure}
  \begin{subfigure}[t]{0.5\textwidth}
    \centering
    \includegraphics[width=\linewidth]{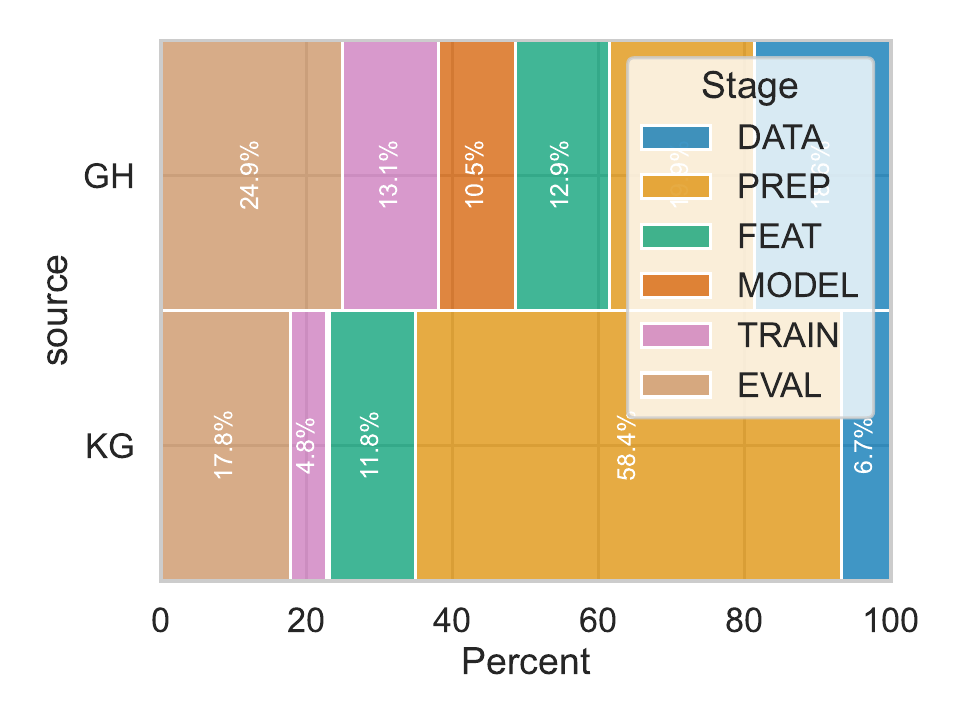}
    \caption{Pipeline stage by source, normalized within source.}
    \label{fig:distribution-stage}
  \end{subfigure}
  \begin{subfigure}[t]{0.5\textwidth}
    \centering
    \includegraphics[width=\linewidth]{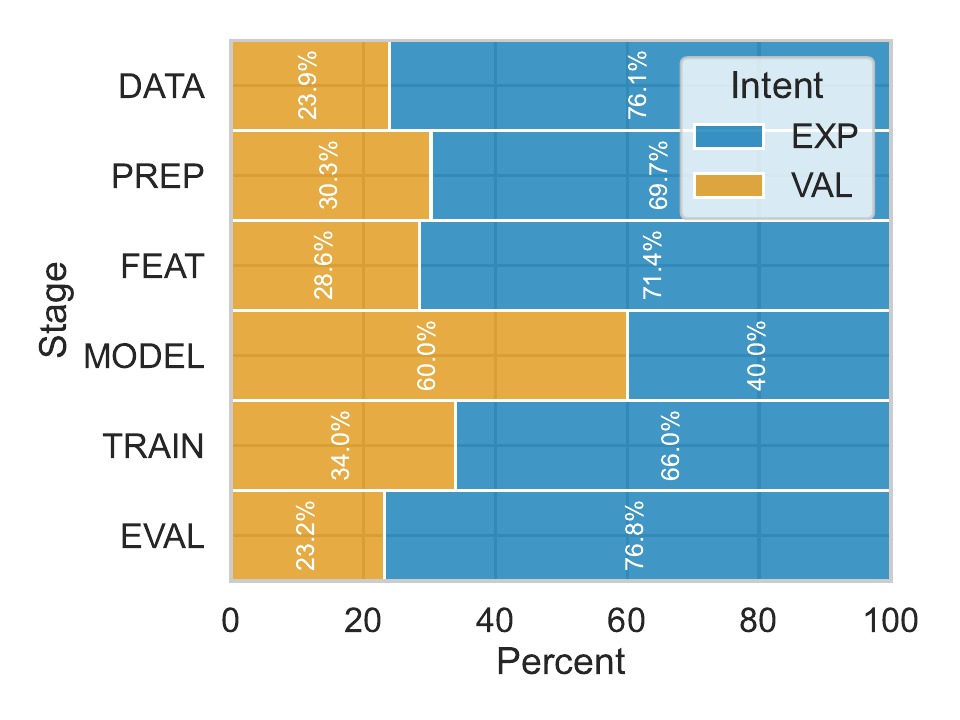}
    \caption{Intent by stage on GH, normalized within stage.}
    \label{fig:distribution-stage-intent}
  \end{subfigure}
  \caption{Distribution of feedback statements across pipeline stages.
  Kaggle concentrates feedback in the PREP stage and barely touches the MODEL stage,
  while GitHub spreads feedback more evenly across stages.
  On GitHub, the share of VAL statements within each stage
  is highest in the MODEL, TRAIN, and PREP stages.}
\end{figure}

This section examines where in the ML pipeline
practitioners author feedback statements,
and how the choice of intent and its subtypes varies by stage.
We organize the analysis as a sequence of three cross-tabulations of increasing granularity.
First, we examine stage by source
to characterize platform-level differences in where feedback appears.
Next, we examine intent by stage,
stratified by source to control for the platform-level effect established in RQ2.
Finally, we examine intent type by stage on the GitHub subset only,
because Kaggle's heavy concentration in data preparation
and its small VAL count
prevent meaningful comparisons.
Two statements (0.25\%) receive a stage code of UNC
because their context did not permit unambiguous assignment,
and we exclude them from the analysis.
This leaves $n=814$ statements across six stages.

\textbf{Notebook feedback is concentrated upstream of model training, not downstream.}
PREP (40\%) is the single largest stage,
followed by EVAL (21\%), DATA (12\%), FEAT (12\%), TRAIN (9\%) and MODEL (6\%).

Based on the results of RQ2
which establishes that the type of intent varies by platform,
we hypothesise that the ML pipeline stage in which feedback statements are used
also differ by platform.
A $\chi^2$ test confirms our hypothesis,
indicating a significant and large association
($\chi^2(5) = 158.04$, $p<0.001$, Cram\'er's $V=0.441$),
and is visualized in Figure~\ref{fig:distribution-stage}.
Feedback in Kaggle notebooks
concentrates predominantly on the PREP stage (58\%),
while feedback in GitHub notebooks
distribute more evenly across DATA (19\%), EVAL (25\%) and TRAIN (13\%).
The MODEL stage is almost entirely absent on Kaggle (0.5\% vs.\ 11\% in GitHub).
This pattern is consistent with the competition format,
which concentrate on data preparation pipelines
and typically consume pre-configured model APIs
rather than defining model architectures from scratch.

To examine whether intent is associated with pipeline stage,
we stratify the analysis by source
since the prior tests establish that it has an effect on both variables.
We restrict further analysis to GitHub statements ($n=381$),
since Kaggle statements ($n=433$) contain only four VAL statements
which makes the effect size estimate unstable.

We hypothesise that assertions are less common in the exploratory ML stages,
since practitioners do not yet know what to test,
and the purpose of these stages is to explore and identify.
A $\chi^2$ test reveals a statistically significant association
($\chi^2(5)=20.34$, $p<0.001$, Cram\'er's $V=0.231$)
and the direction of the association matches our prediction.
Figure~\ref{fig:distribution-stage-intent} shows that
MODEL is the most validation-heavy stage (60\%),
followed by TRAIN (34\%) and FEAT (29\%),
while DATA and EVAL are more exploratory.
\textbf{This indicates that practitioners validate when the code defines reusable structure,}
such as model definitions and training loops,
and inspect when the code is exploratory and data-facing,
such as preprocessing and evaluation.

\begin{figure}
  \begin{center}
    \includegraphics[width=\textwidth]{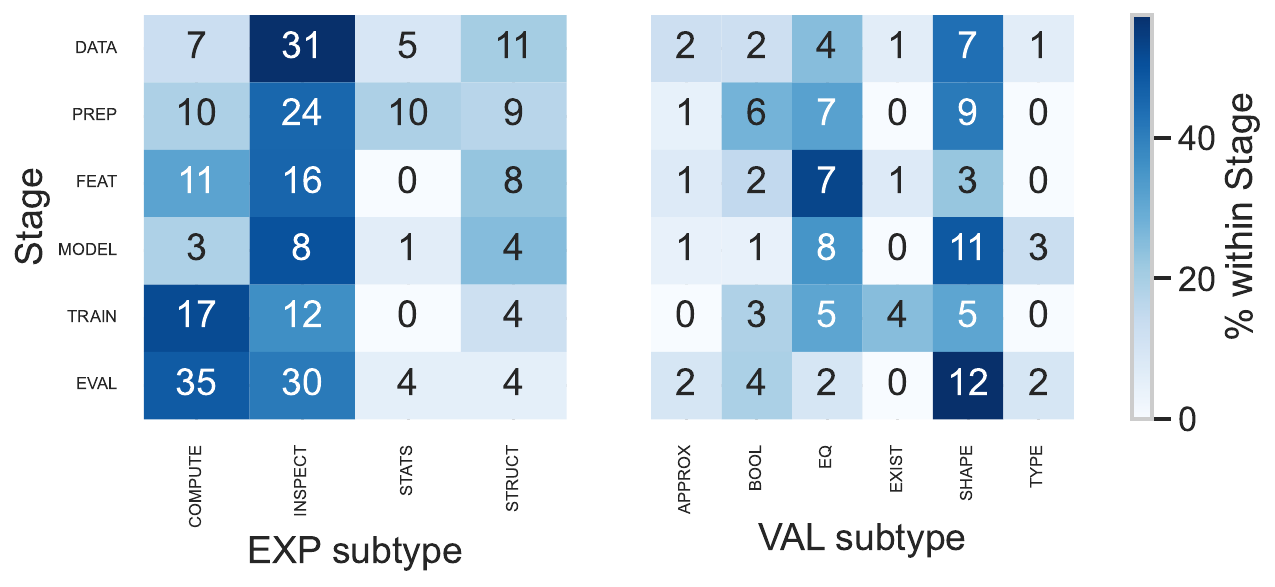}
  \end{center}
  \caption{Intent type distribution by stage on GitHub statements.
  The cells show the count, while the colorbar shows the percentage
  normalized within stage.
  Within EXP statements,
  COMPUTE peaks in TRAIN and EVAL,
  STATS in PREP and DATA,
  and INSPECT dominates every stage.
  Within VAL statements,
  SHAPE is the modal type in four of six stages,
  and EQ appears at every stage.}
  \label{fig:distribution-stage-type}
\end{figure}

Figure~\ref{fig:distribution-stage-type}
shows the distribution of the intent subtypes by stage, for GitHub statements.
Within the GitHub EXP statements ($n=264$),
INSPECT is the dominant exploratory type
and accounts for 36\% to 57\% of the statements within each stage.
COMPUTE peaks in TRAIN (52\%) and EVAL (47\%),
where practitioners compute losses, predictions, and metrics inline.
STATS appears almost exclusively in PREP (19\%) and DATA (9\%),
which matches the role of summary statistics in early data understanding.
STRUCT is small but spread across DATA (20\%), FEAT (23\%), MODEL (25\%), and PREP (17\%).

Within the GitHub VAL statements ($n=117$),
\textbf{the choice of validation type is largely stage-agnostic.}
SHAPE is the modal validation type in four of the six stages
(DATA, EVAL, MODEL, PREP)
and accounts for 39\% to 55\% of the statements within those stages.
EQ is modal in FEAT (50\%) and TRAIN (29\%),
and BOOL appears at every stage in small numbers (4\% to 26\%).
The rarer subtypes (APPROX, EXIST, TYPE) appear too few times
to support stage-level claims.

\subsection{RQ4: Relationship to Known Crash Modes}
\label{sec:rq4}

\begin{table}
  \centering
  \caption{Mapping of VAL subtypes to crash types reported by Wang et al.~\cite{wang2025why}.
  Each row identifies the crash types that the corresponding VAL subtype is designed to pre-empt.}
  \label{tab:wang-mapping}
  \begin{tabularx}{\textwidth}{l X}
    \toprule
    \textbf{VAL subtype} & \textbf{Wang et al.\ crash types} \\
    \midrule
    SHAPE  & Tensor Shape Mismatch, Unsupported Broadcast\\
    EQ     & Data Value Violation, Invalid Argument, Model Initialization Error \\
    BOOL   & Data Value Violation, Invalid Argument \\
    EXIST  & Variable Not Found, Key Error, IO Error \\
    TYPE   & Type Error, Attribute Error \\
    APPROX & (no direct counterpart) \\
    \bottomrule
  \end{tabularx}
\end{table}

RQ1 and RQ2 characterize what feedback statements practitioners author,
and RQ3 shows where they author them.
RQ4 situates those findings against the failure modes
identified by analyzing crashes in Jupyter notebooks.
Wang et al.~\cite{wang2025why} examined ML Jupyter notebooks
collected from GitHub and Kaggle,
and cataloged \emph{crashes}
which halt execution and surface an exception in the cell output.
While the Wang taxonomy identifies the failure mode,
our taxonomy provides a complementary view,
by identifying validation statements
designed to pre-emptively catch such errors.
For each VAL subtype,
we check whether an assertion would
either prevent the crash from occurring
or surface an upstream defect that would otherwise trigger the crash later.
The resulting mapping is shown in Table~\ref{tab:wang-mapping}.

\textbf{VAL-SHAPE} assertions pre-empt
Tensor Shape Mismatch and Unsupported Broadcast crashes,
which are among the most common ML-specific crash types.
Our results are consistent with this prevalence,
as VAL-SHAPE is the most prominent validation type in our corpus,
and accounts for 39\% of VAL statements.

\textbf{VAL-EQ and VAL-BOOL} pre-empt crashes that arise
when downstream code receives semantically incorrect inputs.
Wang et al.\ report that API misuse alone accounts for
66\% of Invalid Argument crashes and 31\% of Data Value Violation crashes.
In our corpus,
architectural constraint assertions (Listing~\ref{lst:4})
pre-empt Model Initialization Error,
while data quality checks (Listing~\ref{lst:54})
pre-empt Data Value Violation crashes,
mirroring the data confusion root cause identified by Wang et al.

\textbf{VAL-EXIST} catches a subset of Variable Not Found and IO Error crashes.
Wang et al.\ attribute most Variable Not Found crashes
to notebook-specific issues such as out-of-order execution.
The remaining cases arise from references to objects
that should have been constructed earlier,
and from missing files or keys.
VAL-EXIST statements that check for the presence of a key, attribute, or filename
before the dependent operation runs,
surface the missing dependency as an explicit assertion failure
rather than a cryptic downstream error.

\textbf{VAL-TYPE} statements enforce the type contract that APIs expect,
and pre-empt the API-misuse subset of Type Error (42\%) and Attribute Error (44\%)
crashes reported by Wang et al.

\textbf{Crashes and validations cover different parts of the failure space.}
Wang et al.\ also report
Out of Memory, Module Not Found, Environment Error, and Request Error crashes,
that fall outside our taxonomy.
Conversely, our taxonomy captures defensive programming practices
that are invisible to a crash-based study.
This is demonstrated by VAL-APPROX statements
that have no direct counterpart in the Wang taxonomy.
Approximate equality assertions verify reproducibility
and catch large drifts in the results
across multiple pipeline runs
and different data versions.
For instance,
Listing~\ref{lst:374} pre-emptively stops the execution
when the NDCG diverges from the expected value.
In the absence of the assertion,
the pipeline would execute and produce an incorrect score,
without raising an exception in the cell output.
Similarly, the training-serving skew detection theme within VAL-EQ (Listing~\ref{lst:118})
catches a divergence that produces no crash signal at all.
The data leakage prevention theme within VAL-EXIST (Listing~\ref{lst:84})
catches a defect that inflates evaluation metrics
without producing any error output.

A study that examines only crashes,
therefore cannot observe these practices
because the practices exist precisely to prevent silent failures
that never reach the cell output as an exception.
The asymmetry suggests that the failure space of ML notebooks
is wider than what crash analysis alone can capture.

\textbf{Our taxonomy of feedback statements
and the Wang et al.\ crash taxonomy,
both converge on data preparation
as the empirical center of gravity
in ML notebook development.}
Wang et al.\ measure where execution fails,
and report that 33\% of crashes occur during data preparation.
We measure where practitioners author feedback statements,
and independently report
that 40\% of feedback statements occur during the same stage.
Despite the inverted lenses,
the convergence of these two independent studies,
suggests that data preparation stage
demands the most attention and effort from practitioners to handle crashes,
and consequently the largest share of tooling investment.

The platform asymmetry from RQ2 and RQ3 is also consistent across both studies.
Wang et al.\ report that GitHub notebooks contain four times more crashes than Kaggle (11\% vs.\ 3\%),
and that the crash distribution across pipeline stages
diverges between the two platforms,
which is similar to what we observe for feedback.

\section{Discussion}

\textbf{Implications for tool design.}
Our results from RQ2 (Section~\ref{sec:rq2})
show that 85\% of feedback statements in ML notebooks are exploratory,
which indicates that practitioners read values off the cell output
rather than enforce them programmatically.
This pattern holds on both platforms
but is most extreme on Kaggle,
where formal assertions account for fewer than one in a hundred feedback statements.
The dominant feedback mode in ML notebook development
is therefore visual confirmation, not assertion-based validation.

Breck et al.~\cite{breck2017ml} argue
that training data deserves the same testing discipline as production code,
and propose a rubric for assessing the readiness of an ML pipeline.
The same authors later introduce \emph{TensorFlow Data Validation}~\cite{breck2019data},
a tool that validates each batch of data against a declared schema.
Data validation libraries such as TFDV and Great Expectations~\footnote{https://greatexpectations.io/}
require practitioners to author the schema as a separate artefact.
Our results indicate that
practitioners already author the same checks in their notebooks,
but in an exploratory rather than declarative form.

We see two concrete directions for tool developers
that amplify what practitioners already do.
First,
EXP-STRUCT statements
such as \lstinline{df.shape} and \lstinline{X_train_under.shape}
serve as informal post-condition checkpoints after data transformations.
A static analyser can detect these statements,
record the values they produce on a successful run,
and propose a corresponding shape assertion,
or a shape constraint in a data schema.
The taxonomy proposed in Section~\ref{sec:rq2}
can be used to identify candidate assertions
that were authored by practitioners as structural probes.
Second,
EXP-STATS calls to \lstinline{.describe()}
that follow a scaling or imputation step
encode an expected post-condition (e.g., mean near zero, no missing values).
A tool can extract these expectations from the visible output
and emit an explicit precondition for the next cell,
or contribute the same expectation
to a Great Expectations suite attached to the notebook.

\begin{lstlisting}[caption={A last-cell boolean expression
that verifies no rows were lost during a transformation.}, label={lst:480}]
test_raw.shape[0] == test.shape[0]
\end{lstlisting}

\textbf{Exploration versus validation}.
Our taxonomy of feedack statements from RQ1 (Section~\ref{sec:rq1})
reveal last-cell statements within the EXP-COMPUTE category,
that are structurally exploratory
but functionally equivalent to assertions.
For instance, Listing~\ref{lst:480} verifies that no rows were lost during a transformation.
Similarly, Listing~\ref{lst:491} verifies that all expected columns are present after one-hot encoding.

\begin{lstlisting}[caption={Verify that all expected columns are present after one-hot encoding.}, label={lst:491}]
pd.Series(options_short_list).isin(dummies.columns).all()
\end{lstlisting}

Since neither expression is wrapped in an \lstinline{assert},
a \lstinline{False} result is visible in the cell output
but does not halt execution.
Consequently, downstream cells in the notebook
continue to run on data
that violates an assumption
that the practitioner authored and enforced manually.
The decision to proceed despite the violation
is therefore implicit and undocumented,
and accumulates technical debt
as data, personnel, and execution environments change
over the lifetime of the notebook~\cite{sculley2015hidden,sambasivan2021everyone,pimentel2019large-scale}.
Promoting the same expression to an \lstinline{assert}
makes the assumption explicit
and turns an undocumented decision into a checked invariant.

The promotion also addresses a reproducibility threat
that is specific to the notebook format.
Notebook cells can execute in any order,
which can leave the kernel in a state
that does not reflect the source code~\cite{wang2020assessing,pimentel2019large-scale}.
A variable assigned in an earlier cell can be silently overwritten,
shadowed by a stale value,
or computed from inputs
that the source code no longer reflects.
A last-cell boolean expression
evaluates against the kernel state at the moment it is first executed,
but offers no protection
when the kernel state diverges from the source
due to an out-of-order execution of the surrounding cells.
An \lstinline{assert} authored at the same point
re-evaluates the condition every time the cell runs,
and fires when the kernel state
no longer satisfies the invariant the practitioner originally checked.

\textbf{Implications for research.}
Our results from RQ3 (Section~\ref{sec:rq3}) show that
Kaggle competition workflows operate primarily on data exploration,
and delegate evaluation of the model to the platform leaderboard.
GitHub, on the other hand, hosts a broader range of activities that spans the entire ML development lifecycle.
Our findings highlight the difference in the kind of work each platform hosts,
and show that the platforms are not interchangeable data sources.
Future studies of ML developer practice
should take this into account,
since drawing data from a single source
will miss the practices on the other.

\section{Threats to Validity}

\textbf{External Validity}.
Our corpus consists of publicly shared notebooks from GitHub and Kaggle.
This excludes notebooks authored privately or as part of proprietary workflows,
where production handoffs, code review, or compliance requirements
may impose stricter validation discipline.
Our findings reflect the feedback practices visible in public notebooks
and future work should focus on industrial ML prototype notebooks
to assess how far our findings extend.

The GitHub data is collected through GitHub's code search on a single date in June 2023.
The Kaggle data is drawn from KGTorrent,
which was collected between November--December 2021.
Notebook practices evolve as libraries, tooling, and platform features change,
and patterns observed in our corpus
may not reflect more recent conventions,
such as agentic coding.
We mitigate this threat
by focusing on patterns rather than specific API usage.

\textbf{Internal Validity}.
Our extraction of last-cell expressions
assumes that the source code in the cell matches the output stored in the notebook.
A practitioner who edits a cell after execution,
without re-running it,
leaves the cell in an inconsistent state.
The expression we extract
can then differ
from the expression that produced the output.
We mitigate this threat
by analysing the static intent of the expression
rather than its runtime value.
Our taxonomy is thus anchored in what the practitioner authored,
not what was last executed.
The threat is therefore confined
to cases where the static intent itself is ambiguous,
which the codebook resolution process is designed to handle.

Our analysis depends on manual coding of feedback statements,
which introduces subjectivity
and the risk of inconsistent application of codes.
We mitigate this threat through the iterative codebook development
and the double-coding procedure described in Section~\ref{sec:method}.
A complete double-coding of the corpus
is the most rigorous remedy
and is left to future replication work.



\textbf{Construct Validity}.
We define feedback
as a combination of assertion statements,
and last-cell expressions that produce visible output.
This excludes feedback delivered through other channels
such as inline print statements that are not in the last position,
logging calls,
and visualisations rendered through magic commands~\footnote{Built-in
magic commands in IPython: https://ipython.readthedocs.io/en/stable/interactive/magics.html}.
Our choice is grounded in prior work that establish the practitioner workflow.
Last-cell expressions are the canonical idiom for inspecting state in Jupyter~\cite{kery2018story,head2019managing}
and assertions are the canonical idiom for enforcing invariants~\cite{kochhar2017revisiting}.
This scope may nevertheless under-represent practitioners
who prefer print-based or log-based feedback.
Our corpus does not exhaust the space of feedback in ML notebooks,
but it captures the two dominant idioms.

The mapping in Section~\ref{sec:rq4}
is established on functional grounds,
by judging whether an assertion would prevent
or surface a given crash type.
The judgement is informed by the Wang root cause analysis
but is itself a construction of this study.
Different judgements are possible at the boundaries,
particularly for the VAL-EXIST to Variable Not Found mapping
which depends on whether the missing reference is a within-cell defect
or a notebook-semantic defect.
We document the mapping rule explicitly
and welcome others to audit or revise it.


\section{Related Work}

\textbf{Bugs and Crashes in ML Notebooks}.
Wang et al.~\cite{wang2025why} catalog 92,542 crashes across 64,031 ML notebooks
collected from GitHub and Kaggle,
and classify the crashes based on their
type, root cause, ML pipeline stage, and the ML library responsible.
De Santana et al.~\cite{desantana2024bug} study Stack Overflow posts,
GitHub commits and interviews with developers to identify high-level issues
when working with notebooks.
Jiang et al.~\cite{jiang2025exploring} extend this taxonomy,
by conducting an empirical study of bugs and vulnerabilities in the source code.
Our study, complements this work
by focusing on asserts and last-cell statements
that practitioners write
to confirm that no failure surfaces in the first place.

A separate line of work targets specific defect classes in notebooks.
Patra and Pradel~\cite{patra2022nalin}
present \textsc{Nalin},
a technique that detects name-value inconsistencies,
where a variable's name suggests one type or shape
but its runtime value carries another.
\textsc{Nalin} is a runtime detection technique,
that targets the same class of latent defects
caught by the VAL-TYPE assertions,
identified in this study.

\textbf{Mining and Analysis of Computational Notebooks}.
Large-scale mining of Jupyter notebooks has become an established methodology.
Pimentel et al.~\cite{pimentel2019large-scale} mine 1.4 million notebooks from GitHub
to study quality and reproducibility.
Psallidas et al.~\cite{psallidas2019data} analyse six million notebooks,
two million enterprise data science pipelines,
and metadata from twelve major data science libraries
to characterise the evolving landscape of computational notebook usage.
Quaranta et al.~\cite{quaranta2021kgtorrent} construct \textsc{KGTorrent},
a public corpus of approximately 250,000 Python notebooks from Kaggle
together with a relational database of metadata.
Ghahfarokhi et al.~\cite{ghahfarokhi2024distilkaggle} extend this foundation,
and present \textsc{DistilKaggle},
a curated dataset of Kaggle notebooks
with finer-grained metadata than \textsc{KGTorrent}.
Grotov et al.~\cite{grotov2022large-scale} compare
code quality between Jupyter notebooks and Python scripts on GitHub.

A complementary methodological contribution comes from Jiang et al.~\cite{jiang2022elevating}
who label notebook cells with their position in the ML development lifecycle
through static data flow analysis.
Their stage taxonomy informs the pipeline stage dimension we use in our coding.
We adopt a similar set of stages,
but apply them at the statement level rather than the cell level
and assign stage through manual coding rather than static analysis.

\textbf{Validation and Testing Practice}.
Empirical work on testing and validation practice in software engineering
has examined how developers use assertions and lightweight checks.
Kochhar et al.~\cite{kochhar2017revisiting} replicate and extend a prior study
of assertion usage on 185 Java projects from GitHub,
and report that assertion adoption remains low despite established benefits.
Vidoni~\cite{vidoni2021evaluating} studies unit testing practices in 177 R packages
and identifies systematic gaps in test coverage
together with sources of testing technical debt.
Both studies establish a baseline,
and report that programmatic validation is under-adopted across language ecosystems.
Our study extends this baseline to the ML notebook setting,
and finds that formal assertions account for 31\% of feedback statements
on GitHub and under 1\% on Kaggle.

In the ML domain,
two prior taxonomies bear directly on our work.
Morovati et al.~\cite{morovati2023bugs} present \textsc{defect4ML},
a benchmark of 100 reproducible bugs
focused on TensorFlow and Keras.
The benchmark is constructed from GitHub and Stack Overflow reports,
and serves as a fault-injection target for downstream tooling research.
Humbatova et al.~\cite{humbatova2020taxonomy} construct a comprehensive taxonomy of faults
in deep learning systems
by analyzing 1059 artefacts from GitHub and Stack Overflow,
and structured interviews with 20 practitioners.
While both taxonomies catalogue defects,
our taxonomy catalogues the practices designed to catch them.

The closest empirical study of validation in ML projects
is Openja et al.~\cite{openja2024empirical},
who investigate the test code present in eleven ML projects across four application domains.
Their study examines the formal test suites that ML projects ship
and reports that test coverage and structure
deviate substantially from conventional software projects.
Our study extends this finding
by examining validation authored within notebooks
rather than in separate test files,
which is the dominant locus of ML development before production handoff.
Additionally, we include exploratory feedback alongside formal validation,
which captures a class of practice that test-suite analysis cannot observe.

\section{Conclusion}

We mine 297,851 notebooks from GitHub and Kaggle,
extract 1,092,780 feedback statements,
and code a sample of 816
to characterize feedback practice in ML Jupyter notebooks.
Our central contribution is a taxonomy of feedback statements,
that distinguishes exploratory feedback from formal validation
and refines each into subtypes grounded in practitioner intent.
The taxonomy reveals that GitHub and Kaggle host qualitatively different modes of ML work,
and that source platform should be treated as a confounder
in empirical studies of ML developer practice.
Mapping the taxonomy onto a crash taxonomy
surfaces silent failures
that crash analysis cannot observe.
Future work should extend the analysis to industrial ML notebooks,
where production handoffs and code review
may impose stricter validation discipline,
and to longitudinal corpora
that capture how feedback practice evolves as notebook tooling matures.

\section{Data Availability}

Our replication package contains
the full dataset of 1M+ feedback statements,
data collection and analysis scripts,
and the final codebook used to derive the taxonomy~\cite{replication}.

\bibliography{bibliography}

\end{document}